\documentclass[prl, twocolumn, preprint, 10pt]{revtex4-2}

\usepackage{subfigure}
\usepackage{xspace}
\usepackage{graphicx} 
\usepackage{gensymb}
\usepackage{amsmath}
\usepackage{subcaption}
\usepackage{esint}
\usepackage{hyperref}

\def\lsim{\mathrel{\raise.3ex\hbox{$<$\kern-.75em\lower1ex\hbox{$\sim$}}}}
\def\gsim{\mathrel{\raise.3ex\hbox{$>$\kern-.75em\lower1ex\hbox{$\sim$}}}}

\def\m87{M87$^*$\xspace}
\def\sgra{Sgr~A$^*$\xspace}

\begin{document}
\title{Constraining Black Hole Parameters from Shadow and Inner-Shadow Morphology Considering Effects of Thick Accretion Disks}

\author{Julien A. Kearns}
\affiliation{Department of Astronomy, University of Virginia, Charlottesville, VA, 22904, USA}

\author{Dominic O. Chang}
\affiliation{Black Hole Initiative At Harvard University, 20 Garden Street, Cambridge, MA 02138, USA}
\affiliation{Center for Astrophysics, Harvard \& Smithsonian, 60 Garden Street, Cambridge, MA 02138, USA}

\author{Daniel C. M. Palumbo}
\affiliation{Black Hole Initiative At Harvard University, 20 Garden Street, Cambridge, MA 02138, USA}
\affiliation{Center for Astrophysics, Harvard \& Smithsonian, 60 Garden Street, Cambridge, MA 02138, USA}

\author{Shane W. Davis}
\affiliation{Department of Astronomy, University of Virginia, Charlottesville, VA, 22904, USA}

\begin{abstract}
    We study the effects of emission geometry on the capability to constrain black hole parameters from measurements of the shadow and inner-shadow of a Reissner-Nordström black hole.
    We investigate constraints on mass, charge, observer inclination, and emission co-latitude made from measurements of these features, where we show that inner shadow asymmetry serves as a robust probe of observer inclination, and that  measurements of inner-shadow size can be used to constrain charge and mass-to-distance ratio.
    However, we find that measurements from the size of the inner-shadow are systematically biased by emission geometry.
    Our results underscore the importance of considering geometric variation when modeling black hole accretion systems. 
    We study the constraining capabilities of observations of \m87-like and \sgra-like systems within the context of the BHEX and ngEHT future observatories.
\end{abstract}

\maketitle
\section{Introduction}
\label{sec:introduction}
The Event Horizon Telescope Collaboration (EHTC) made observations of the centers of the galaxy M87 and the Milky Way to produce the first pictures of the shadows of black holes \m87 and \sgra \cite{m87I,m87IV,EHTC_SgrA_I,EHTC_SgrA_III}.
The images depict ring-like features of ${\sim}40{-}50\;\mu$as in extent that have asymmetric brightness distributions, and are consistent with previous theoretical predictions of the appearances of black holes \citep[][]{Bardeen_1973,Luminet,Heino}. 
These observations allowed the EHTC to establish constraints on the mass of \m87 and the orientation of its spin axis with respect to Earth's line of sight from measurements of the apparent shadow \cite{m87V,m87VI}.

Black hole images are predicted to be composed of emission from multiple nested ring-like features, called photon rings, which converge in shape and size to a critical curve bounding the black hole shadow \citep[][]{MJohnsonRing, Gralla_Lupsasca}.
The shape and morphology of the critical curve is agnostic to the underlying accretion physics, and is instead primarily shaped by black hole mass-to-distance ratio, spin and charge \citep{Johannsen_2010, Walia_2024}.
These features suggest that photon ring measurements could be used to constrain spacetime parameters.
Individual photon rings are, however, unresolvable at the current nominal resolution of the EHT, although space based VLBI missions could discern properties of the first photon ring \citep[][]{Lupsasca_Space, BHEX_Concept}.

Another feature seen in simulated images of black hole accretion flows is the inner-shadow.
The inner-shadow is associated with either the footprint of the black hole jet or accretion disk boundary on the event horizon.
Previous authors have proposed the inner-shadow as an additional probe of spacetime properties \cite{Chael_2021}, which could be characterized from observations in the near future \citep[][]{ngehtarray,ngEHT_Concept}.
However, unlike the shadow, the appearance of the inner-shadow is dependent on accretion properties that affects the jet opening angle and disk scale height \citep[see, for example, the ray traced accretion images in the appendix of][]{m87V}. 
Thus characterizing the dependency of the inner-shadow morphology on both spacetime and accretion properties will be crucial to interpreting measurements in the near future.

This work focuses on constraints that can be derived from measurements of the shadow and inner-shadow around black holes, and how such measurements can be confounded by variations in emission geometries.
We focus on the case of a charged, spherically symmetric black hole, but expect the effects of charge on the size of the image features to be qualitatively similar to effects caused by spin. 

The outline of this paper is as follows:
In the first section, we explain our ray tracing procedure. 
In the second section, we use our ray tracing procedure to study the shadow and inner-shadow image features of accretion flows around black holes, and how they can be used to constrain various spacetime parameters.
We specifically study scenarios that are relevant to observations of \m87 and \sgra.
Finally, we conclude by summarizing our results.

\begin{figure}
    \includegraphics[width=0.475\textwidth]{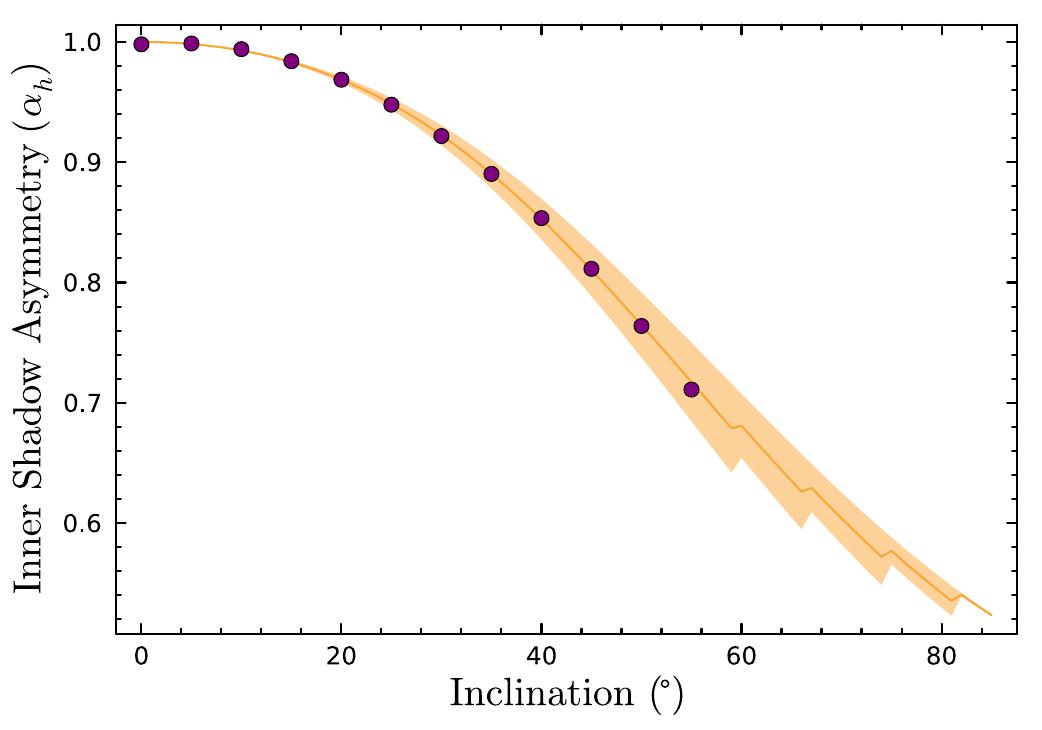}
    \caption{\label{fig:inner_shadow_morphology}$\alpha_h$ vs. $i$, for a Reissner-Nordström black hole over a range of charges $q\in[0,1]$ and emission co-latitudes $s\in[0\degree,30\degree]$. 
    We find a best fit quadratic curve $\alpha_h(i)=-1\times 10^{-4}(i-3.1\degree)^2+1$ that is traced with purple points for values of $i\in[0\degree,50\degree]$.
    The orange region represents the range of inner-shadow asymmetry values that are consistent with each inclination parameter. The thinness of the orange region implies a strong relationship between $\alpha_h$ and $i$. 
    }
\end{figure}

\section{Ray Tracing Method}

\label{sec:raytracing}

Here, we summarize our method for ray tracing surfaces in static, spherically symmetric spacetimes where we take our observer to be at $r{=}\infty$. 

\subsection{Null Geodesics in Spherically Symmetric Black Hole Spacetimes}

We first begin with the metric ansatz,
\begin{align}
    ds^2 = -f(r)\,c^2\dot t^2 +\frac{\dot r^2}{f(r)} +r^2[\dot\theta^2 + \sin^2\theta \,\dot\phi^2],\label{eqn:line-element}
\end{align}
where $t$ is the time coordinate of the asymptotic observer, $r$ is the areal radius of the spacetime, $\theta$ and $\phi$ are the spherical inclination and azimuthal angles.
We will take $G{=}M{=}c{=}1$ hereafter.
We use $\dot{x}$ to indicate a derivative of $x$ with respect to the affine parameter $\lambda$.
The time translation and rotational symmetries of \autoref{eqn:line-element} imply that geodesics are characterized in terms of conserved energy,
\begin{align}
    E 
        &=f(r)\,\dot t,
\end{align} 
and azimuthal angular momentum,
\begin{align}
    L 
        &=r^2\sin^2\theta\,\dot\phi.
\end{align} 

We first consider geodesic motion in the equatorial plane. 
The line element of \autoref{eqn:line-element} then becomes
\begin{equation}
    \frac{d\phi}{d r} =  \pm\frac{b}{\sqrt{R(r)}} ,
    \label{eqn:radial-potential}
\end{equation}
where we have defined the radial potential,
\begin{align}
     R(r)=\sqrt{r^2\left[r^2 -f(r)\frac{L^2}{E^2}\right]}.\label{eqn:radial_potential}
\end{align}

The winding angle, $\psi$, accrued by a photon trajectory can then be written as
\begin{align}
    \psi(r_s)=\Delta\phi
        &=\fint_{r_s}^{\infty}\pm\frac{b}{\sqrt{R(r)}}\,dr,\label{eqn:phi_int}
\end{align}
where $\fint$ indicates a path dependent integral, $r_s$ is the radius of emission, $b{=}L/E$ is the impact parameter of the photon, and $\pm$ indicates whether the radial momentum of the photon is outgoing or ingoing.

Alternatively, we can use the Hamilton-Jacobi Equations \citep[][]{Carter} to bring \autoref{eqn:line-element} to the form
\begin{align}
    \frac{r^2\left(E^2-\frac{1}{r^2}\left(\frac{dr}{d\tau}\right)^2\right)}{f(r)}=\left(\frac{d\theta}{d\tau}\right)^2+\frac{L^2}{\sin^2\theta}=Q+L^2,
\end{align}
where $Q$ is the Carter integral, and $d\tau=\frac{1}{r^2}d\lambda$ is the Mino-time.
This expression is used to define the integral equations:
\begin{subequations}
\begin{align}
    \frac{d\tau}{dr} 
        &=\frac{1}{E^2\sqrt{\,r^2\left[r^2-\frac{(Q+L^2)}{E^2}\,f(r)\right]}}\label{eqn:hje_radial},\\
    \frac{d\tau}{d\theta} 
        &=\frac{1}{\sqrt{Q-L^2\cot^2\theta}}.\label{eqn:hje_theta}
\end{align}
\end{subequations}
\autoref{eqn:hje_radial} is of the same form as \autoref{eqn:radial-potential}.
We can therefore directly relate the Mino-time to the winding angle, and write \autoref{eqn:hje_theta} as
\begin{align}
    \psi(\theta_s,i) = \fint_{\theta_s}^{i} \pm \frac{E \,b}{\sqrt{Q-L^2\cot^2\theta}}\,d\theta,\label{eqn:winding_angle_theta}
\end{align}
where now, $b=\sqrt{(Q+L^2)}/E$, $\theta_s$ is the emission inclination, and $i$ is the observer inclination.
It follows that
\begin{align}
    \psi(r_s)=\psi(\theta_s,\theta_0).
\end{align}

In the following, we will specialize to the case of an electrically charged, spherically symmetric black hole with \citep[][]{reissner,weyl,nordstrom},
\begin{align}
    f(r) 
        &=1-\frac{2M}{r} +\frac{M^2q^2}{r^2},
\end{align}
where $q$ is the black hole charge.
This black hole has an outer event horizon at
\begin{align}
    r_h
        &= M\left(1 +\sqrt{1-q^2}\right) ,
\end{align}
where we have restored $M$ to make the mass dependence explicit.
\subsection{Observer's Screen Coordinates}

We use the coordinates of \cite{Bardeen_1973} for the screen of our observer.
The horizontal and vertical coordinates, $(x,y)$, are defined as
\begin{subequations}
\begin{align}
    x
        &=-\frac{1}{r}\frac{\dot\phi}{\dot t}=-\frac{L}{E}\frac{1}{\sin i}, \quad\text{and}\\
    y 
        &=\frac{1}{r}\frac{\dot\theta}{\dot t}=\frac{1}{E}\sqrt{Q-L^2\cot i},
\end{align}
\label{eqn:bardeen}
\end{subequations}
at $r{=}\infty$.
We will also use polar coordinates on the observer's screen with
\begin{subequations}
\begin{align}
    b
        &=\sqrt{x^2+y^2} ,\quad\text{and}\\
    \varphi
        &=\arctan\left(\frac{y}{x}\right).
\end{align}
\label{eqn:polar}
\end{subequations}

\subsection{Ray Tracing the Shadow and Inner-Shadow of an Accretion Disk }

\begin{figure}
  \centering
  \begin{subfigure}
    \centering
    \includegraphics[width=0.49\textwidth]{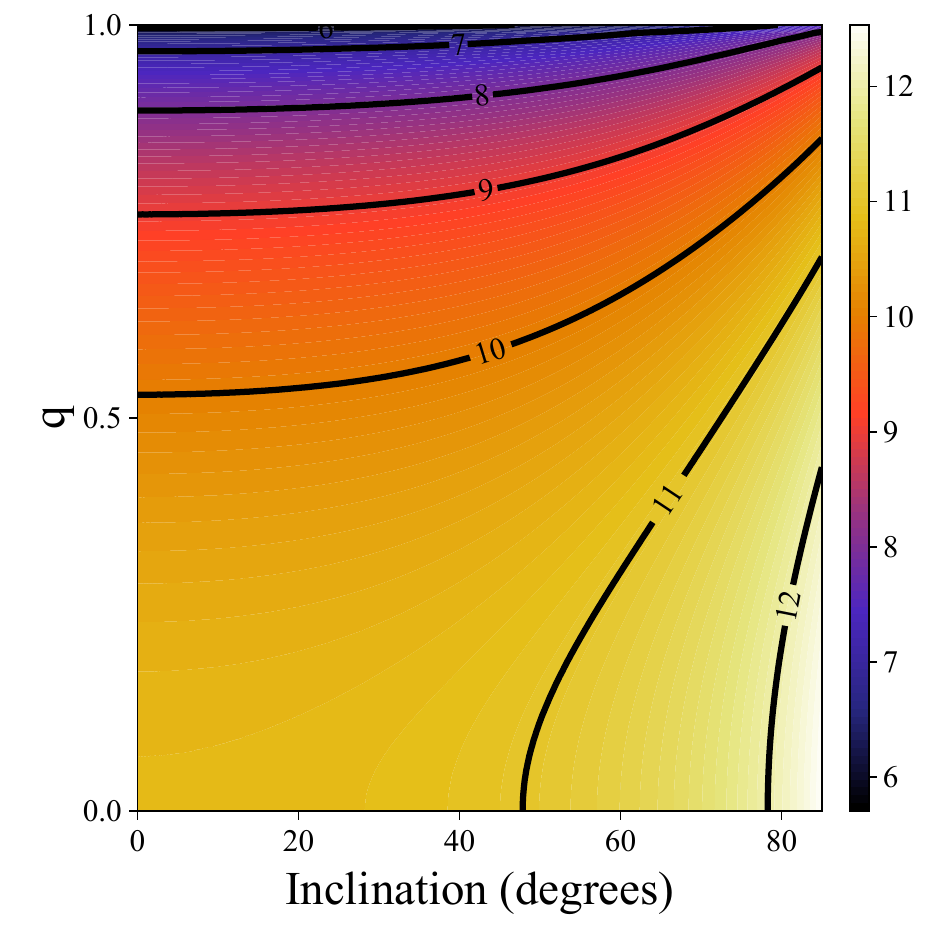}
  \end{subfigure}
  \begin{subfigure}
    \centering
    \includegraphics[width=0.49\textwidth]{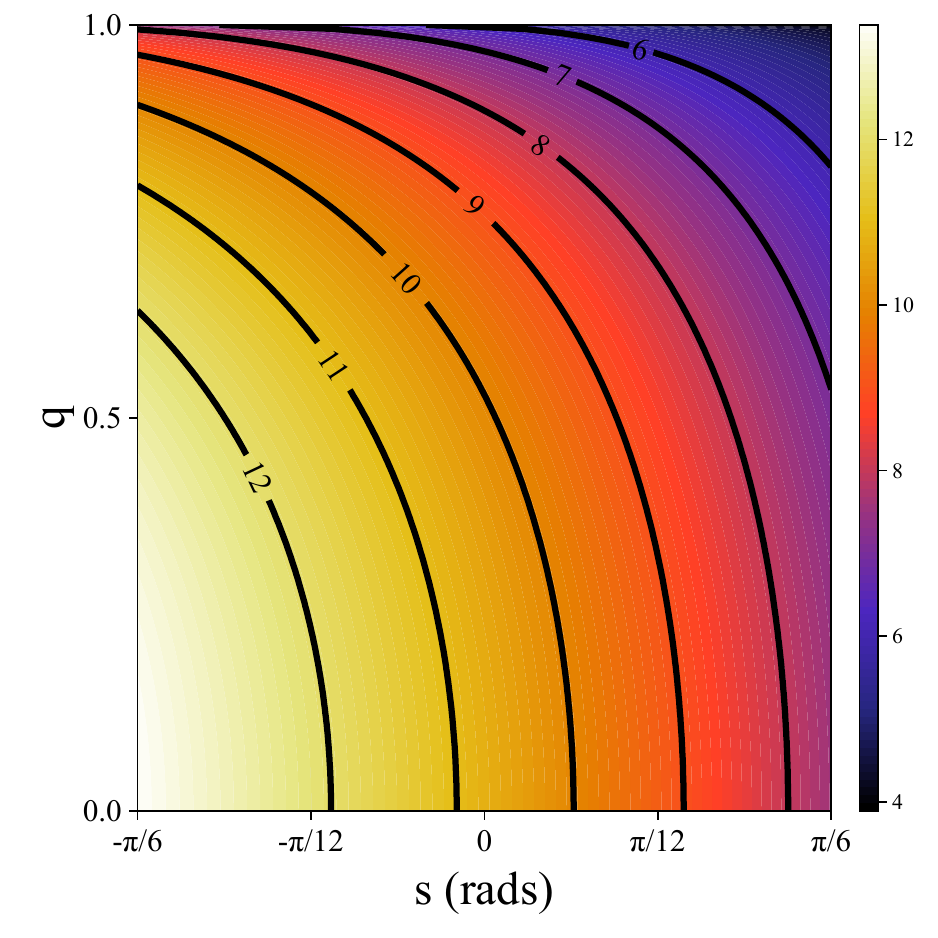}
  \end{subfigure}
  \caption{\label{fig:ivq}(Top) Filled contour plot of all potential $\overline{b_h}$ values across different inclinations and charges.
  Our values are chosen for a black hole with mass-to-distance ratio $M/D=3.78 \;\mu as$ and emission co-latitude $s=0$.
  Isoradial contours of $\overline{b_h}$ at whole number intervals are shown in black.
  (Bottom) Filled contour map of all potential $\overline{b_h}$ values across different emission co-latitudes and charges of a black hole with mass-to-distance ratio $M/D=3.78\;\mu as$ and viewing inclination $i=0$.
  Although only positive co-latitudes are relevant for observations, we plot values of $\overline{b_h}$ for the range $s\in[-30\degree,30\degree]$ to help illustrate general trends.}
\end{figure}

Black hole accretion flows can have differing emission geometries and often feature turbulent structures that can modify their appearance \citep{m87V}.
Despite these complications, all high-resolution images of black holes are expected to have the shadow and the inner-shadow as persistent features \citep{Cunha_Light_Rings,MJohnsonRing,Chael_2021}.
Thus, we will extend the definition of the inner-shadow in \cite{Chael_2021} to include thick emission surfaces with a boundary at an ``emission co-latitude" of $s$ at the horizon.
\subsubsection{The Shadow of the Black Hole}
We define the shadow of a black hole as the region in the image that is bounded by the critical curve \citep[][]{Luminet}. 
Although the critical curve is unlikely to be observed, its approximation in the form of the $n{=}1$ photon ring will be detectable by future space based VLBI missions \citep[][]{BHEX_Concept}.
We will nevertheless work with the idealized assumption that the critical curve can be exactly measured.
The appearance of the critical curve is determined by the impact parameter $b=b_c$ and emission radius $r_c$, that satisfies

\begin{subequations}
\begin{align}
    R(r_c) &=0\\
    \partial_rR(r_c) &=0\\
    \partial^2_r R(r_c) &<0.
\end{align}
\end{subequations}

For the case of a Reissner-Nordström black hole, \autoref{eqn:radial_potential} gives a photon sphere radius of 
\begin{align}
    r_c(q)
        &=\frac{1}{2}\left(3+\sqrt{9-8\,q^2}\right) ,
\end{align}
and screen radius of
\begin{align}
    b_c(q)=\frac{\sqrt{27-36\,q^2+8\,q^4+(9-8\,q^2)^\frac{3}{2}}}{\sqrt{2}\sqrt{1-q^2}} \label{eqn:critical_curve}.
\end{align}

\subsubsection{Inner-Shadow of a Thick Disk}

The inner-shadow corresponds to the image feature where the accretion disk or jet meets the event horizon, and can thus be thought of as the direct image of a co-latitude curve on the horizon.
This feature manifests as a sharp boundary in simulated images of black hole accretion flows where there is no/very little emission seen in its interior.

We generate images of the inner shadow seen by an observer viewing an accretion disk that is tilted at an inclination of $i$ with respect to their line-of-sight.
Taking \autoref{eqn:winding_angle_theta}, \autoref{eqn:bardeen}  and \autoref{eqn:polar} with $\theta_s=\pi/2{-}s$ gives

\begin{align}
\psi_{ss}(\varphi,\theta_s) = 
\begin{cases}
    - \psi(\varphi,\theta_s,i), &\text{if } \psi(\varphi,\theta_s) < 0\\
     \pi - \psi(\varphi,\theta_s), &\text{if } \psi(\varphi,\theta_s) \geq 0,
\end{cases}
\end{align}
with
\begin{align}
    \psi(\varphi,\theta_s)&=\arctan\left[\frac{\cot\theta_s}{\sqrt{1-\cos^2\varphi\csc^2\theta_s\sin^2 i}}\right]\notag\\
    &\qquad-\arctan\left[{\frac{1}{\tan{i}\sin{\varphi}}}\right].
\end{align}

We interpret $s$ as the ``emission co-latitude" of the inner-shadow, and ray trace this boundary by numerically solving the equation,
\begin{align}
    \psi_{ss}(\varphi,\pi/2{-}s,i)  
    =\psi(b_h),
\end{align} 
for the inner-horizon screen radius $b_h$ with the \texttt{Roots.jl} package \citep[][]{Roots.jl}.

\begin{figure}
  \centering
  \begin{subfigure}
    \centering
    \includegraphics[width=0.425\textwidth]{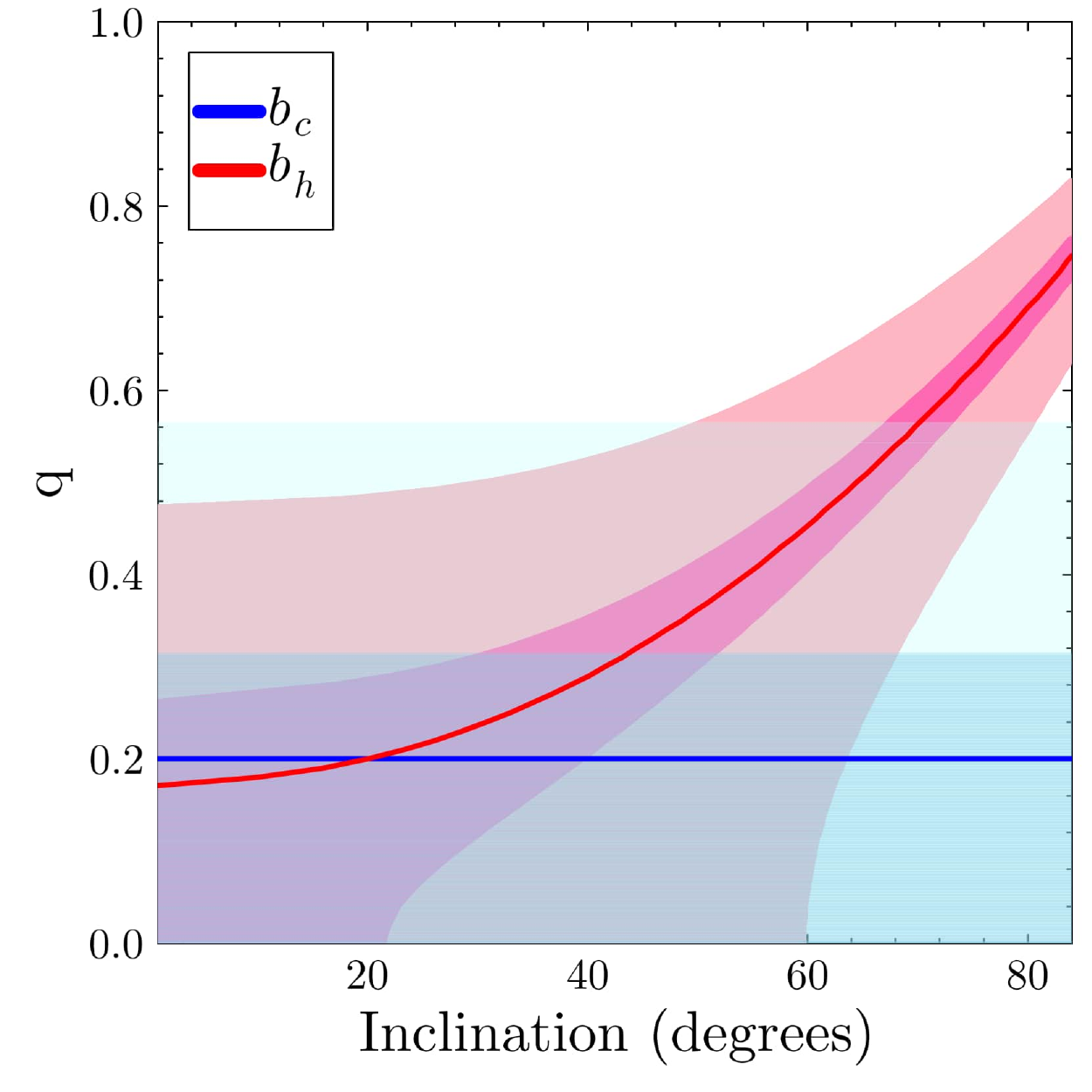}
  \end{subfigure}
  \begin{subfigure}
    \centering
    \includegraphics[width=0.425\textwidth]{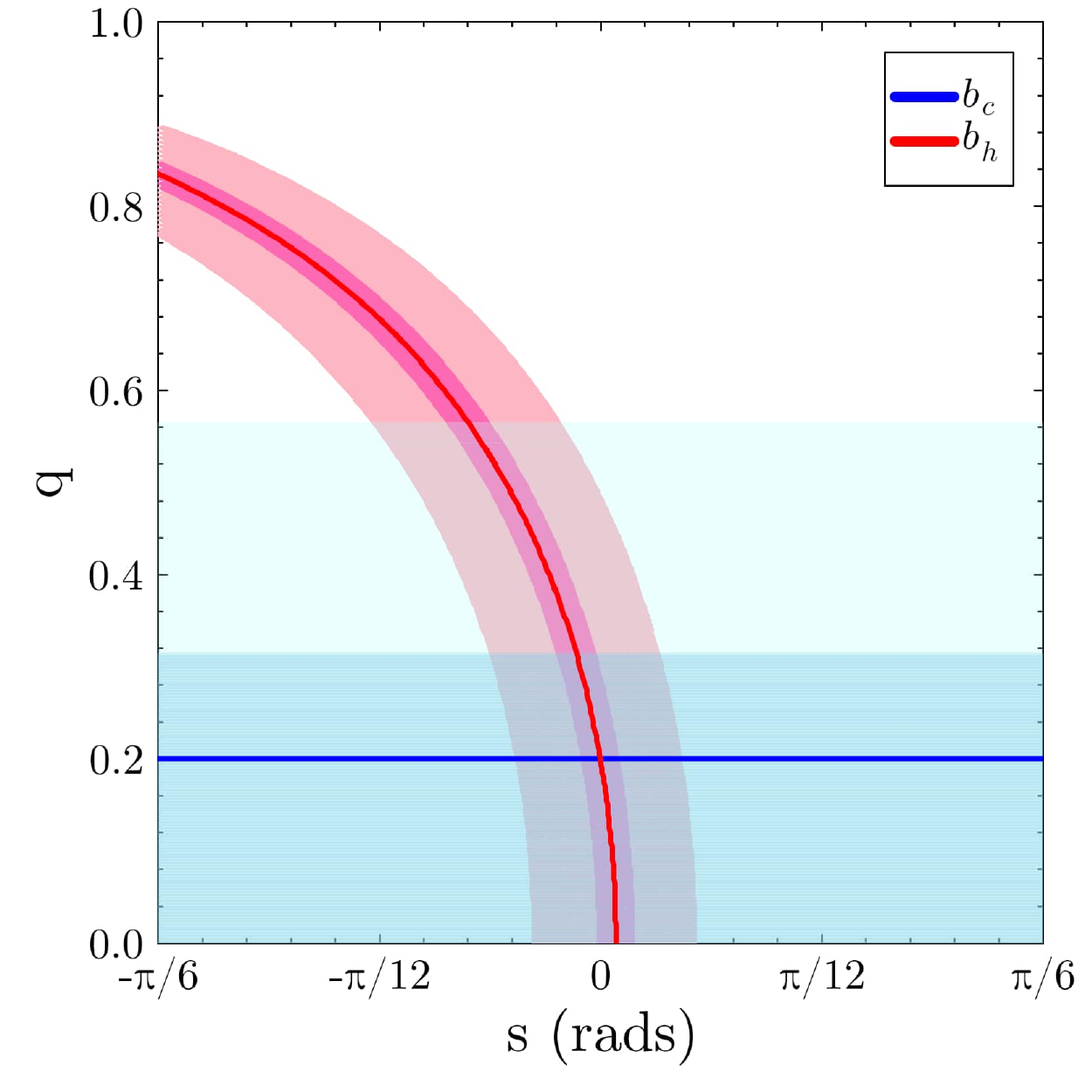}
  \end{subfigure}
    \caption{\label{fig:constraints}(Top) Example of constraints made on black hole viewing inclination $i$, and charge $q$ from measurement of the average radius of the inner-shadow $\overline{b_h}$ (red) and critical curve (blue).  
    The measurement shown is for a black hole with true parameters of $q=0.2$ and $i = 20\degree$. 
    (Bottom) Example of constraints on black hole emission co-latitude $s$, and charge $q$ from measurement of the average radius of the inner-shadow, $\overline{b_h}$. The measurement shown is for a black hole with true parameters of $q=0.2$ and $s = 0\degree$.  
    The thin-dark, and thick-light bands in both panels show the 1\% and 5\% uncertainty regions respectively.}  
    
\end{figure}

\section{Spacetime Constraints from Measurements of the Shadow and Inner-Shadow Features}

\label{sec:results}

In this section, we study the capability for measurements of the size and shape of the shadow and inner-shadow to constrain black hole mass-to-distance ratio $M/D$, charge $q$, observer inclination $i$, and emission co-latitude $s$ between the accretion disk and the horizon as measured in the lab frame.
We also show, through examples, how measurement uncertainties and model assumptions can affect the quality of our constraints.  
We calculate uncertainties that are consistent with the potential resolving power of the future ngEHT and BHEX missions when they will observe \m87 and \sgra \citep{ngEHT_Concept,BHEX_Concept}.
For studies where the known mass is assumed, we use a nominal mass-to-distance ratio of $3.78\;\mu as$, though we note that a change in the mass-to-distance ratio of the black hole is equivalent to applying a uniform scaling to the sizes of the shadow and inner-shadow.

\autoref{eqn:critical_curve} suggests that size of the shadow is completely determined by $q$ and $M/D$, and will always appear circular in shape.
In contrast, the morphology of the inner-shadow will vary with respect to the $M/D$, $q$, $i$, and $s$.
In order to quantify this dependency, we define $x_h$ and $y_h$ as the inner-shadow principal axes, with $x_h$ being the longest horizontal cross-section and $y_h$ the longest vertical.
These quantities are used to define the average size of the inner-shadow as
\begin{align}
    \overline{b_h}
        &=\frac{x_h+y_h}{2}, 
\end{align}
and the  asymmetry, $\alpha_h$, as
\begin{equation}
    \alpha_h=\frac{x_h}{y_h}.
\end{equation}

\begin{figure} 
    \centering 
    \hspace{-0.6cm}
    \includegraphics[width=.5\textwidth]{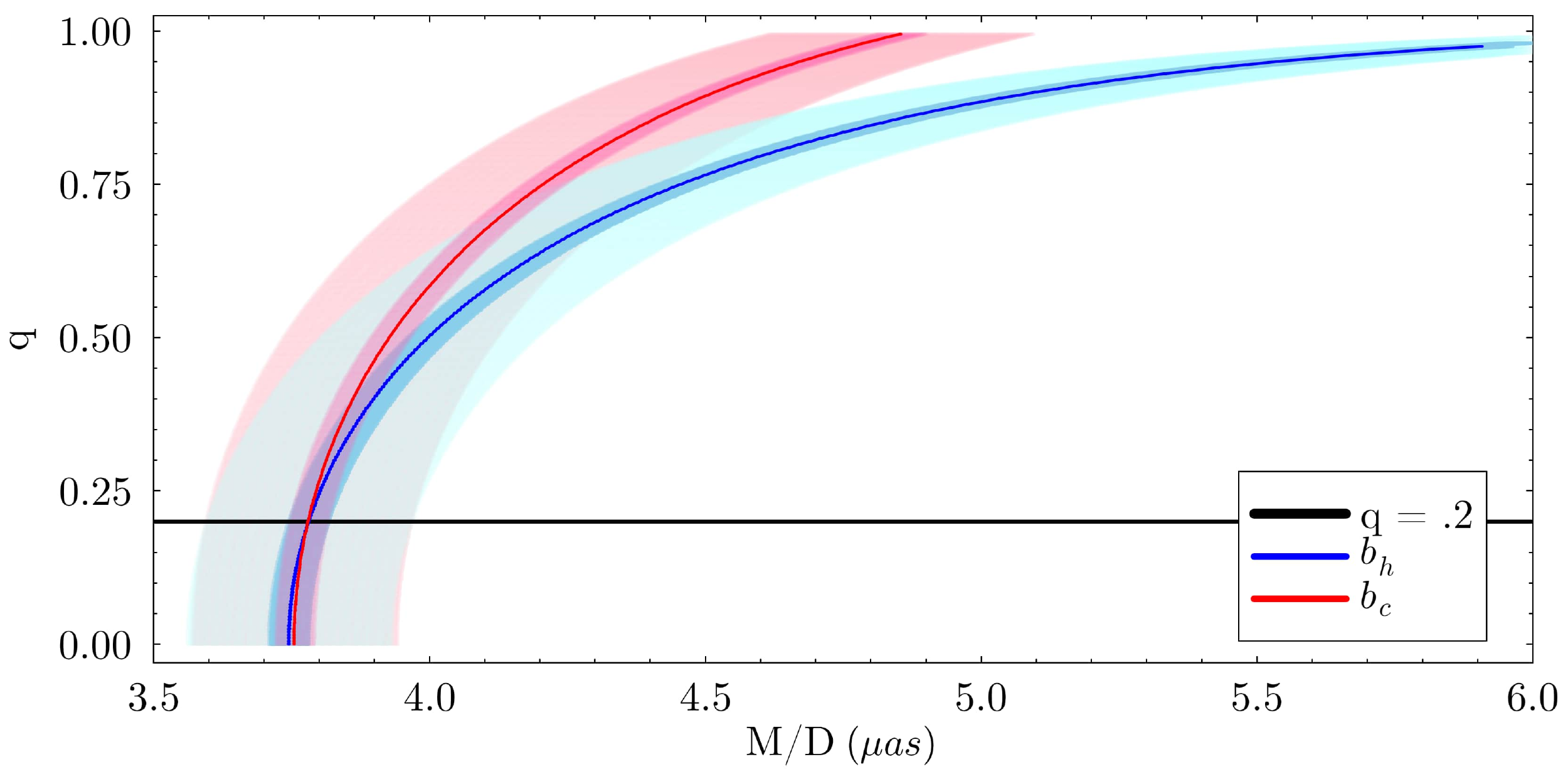} 
    
    \hspace{-0.6cm}
    \includegraphics[width=.5\textwidth]{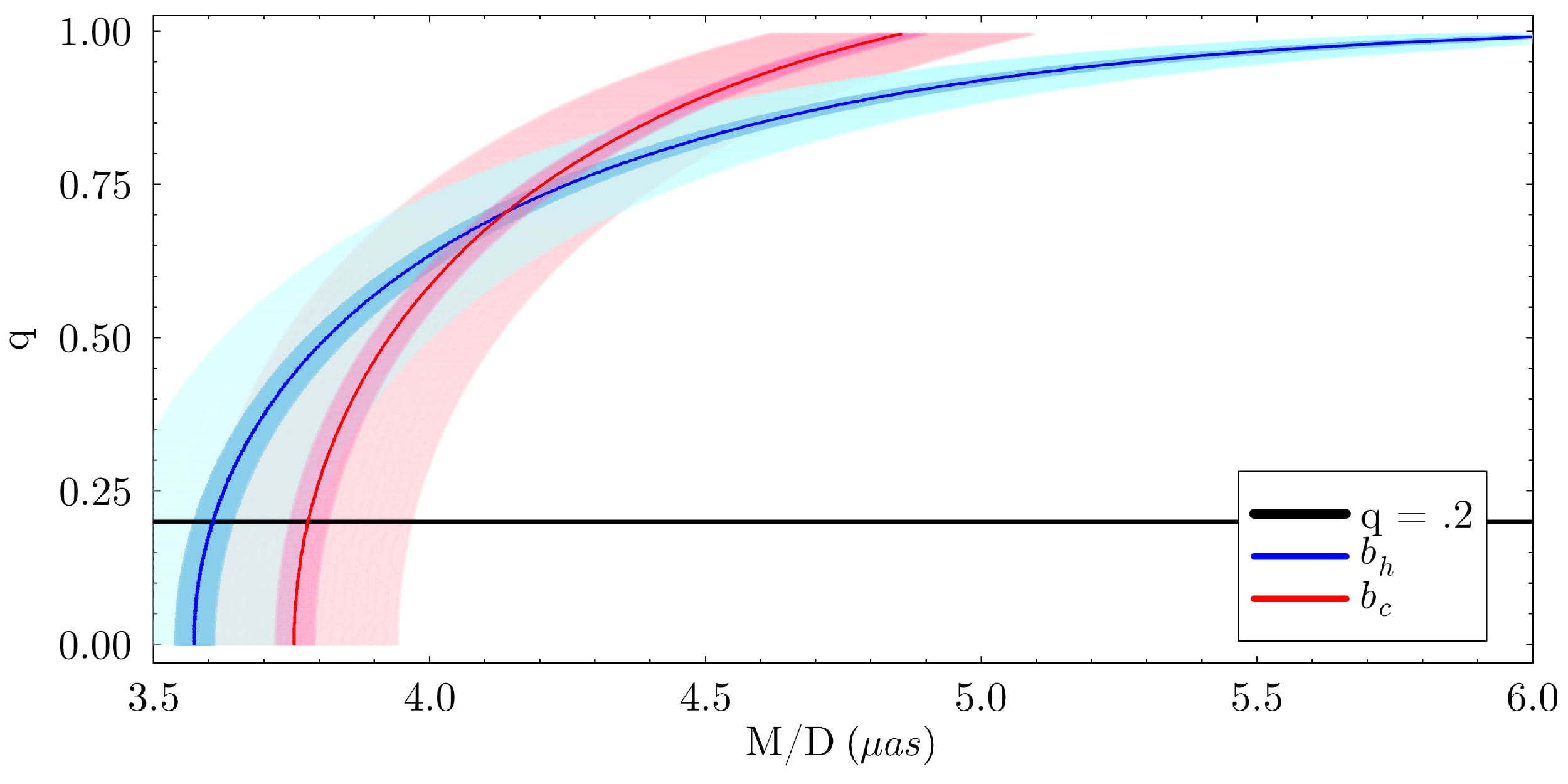} 
    
    \hspace{-0.6cm}
    \includegraphics[width=.5\textwidth]{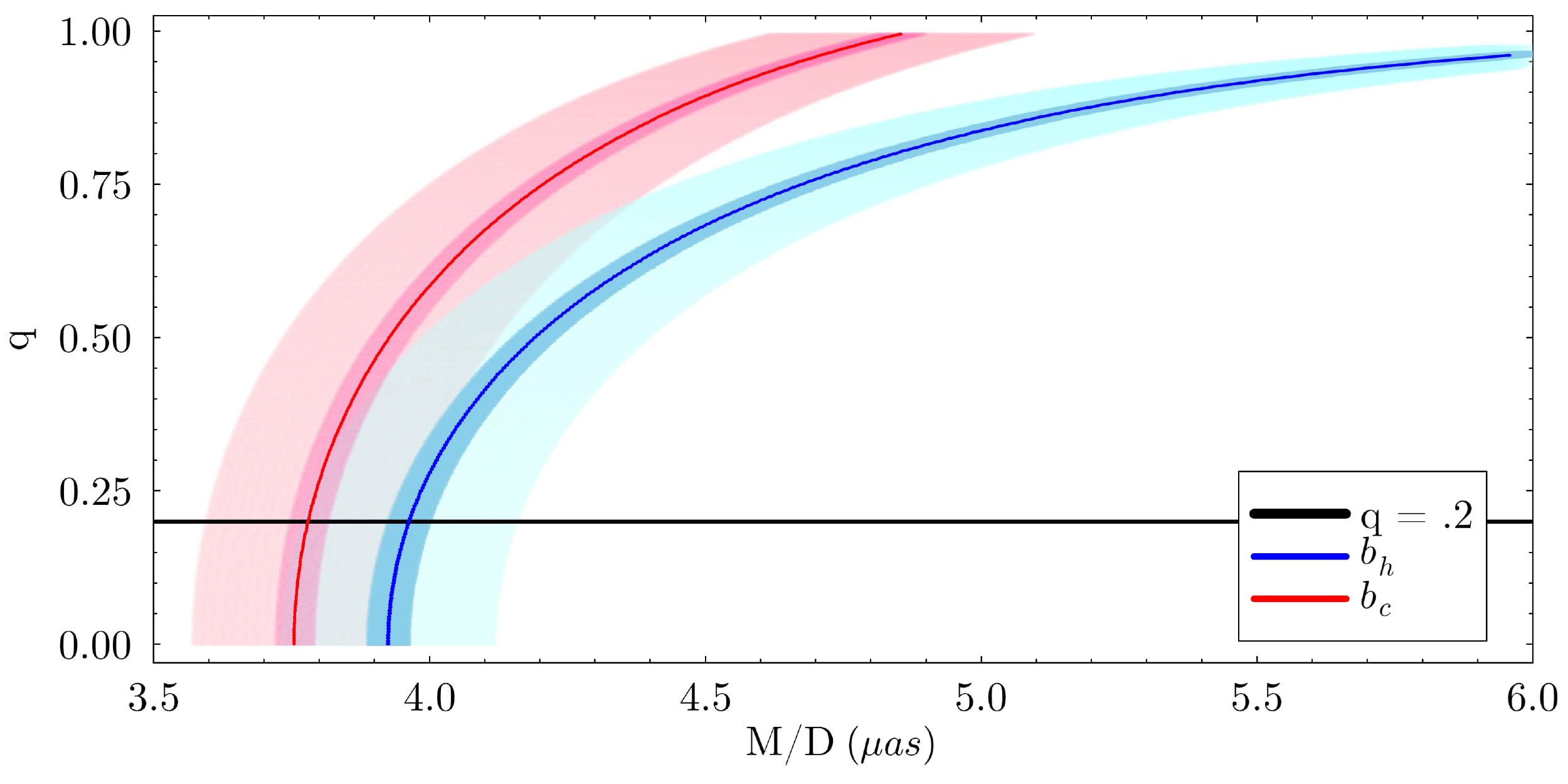} 
    
    \caption{ 
    The effects of assumptions on emission geometry on $M/D$ and $q$ constraints.
    Constraints are made from independent measurements of $\overline{b_h}$ (blue) and $b_c$ (red) for a black hole with true parameters of $q=0.2$, $M/D = 3.78\;\mu as$, and $s=0\degree$. 
    The bands represent regions of measurement uncertainty, with the thinner and thicker bands filling the 1\% and 5\% uncertainty regions respectively.
    The top panel shows the constraints made if the correct emission co-latitude is assumed, $s{=}0\degree$.
    The center and bottom panels shows the attempted constraints made if the emission co-latitude is incorrectly assumed.
    The center panel shows the case where the true emission co-latitude is at $s{=}0\degree$ but is assumed to be $s{=5}\degree$.
    The bottom plot shows the same, but where the truth is $s{=}5\degree$ and is assumed to be $s{=0}\degree$. 
    }
    \label{fig:stacked_plots}
\end{figure}

\subsection{\sgra-like system: Measurement of Charge, Inclination and Scale Height}

We study how measurements of the shadow and inner-shadow morphology could be used to constrain $q$, $i$, and $s$. 
Assuming a precise measurement of the black hole's $M/D$.
We will assume a known $M/D$ of $3.78\;\mu as$, though the main takeaways of our results are generic and can be re-scaled to other values of $M/D$.
This scenario is relevant for observations of systems with a well constrained mass like that of \sgra \citep[see][]{keck_sgra}.

First, we show the relationship between $\alpha_h$ and the observer's inclination in 
\autoref{fig:inner_shadow_morphology}.
The top panel shows a monotonically decreasing trend in $\alpha_h$ with changing $i$, for $q\in[0,1]$ and $s\in[0\degree,30\degree]$.
This trend is very well described by a parabola for values of $i\in[0\degree,50\degree]$ where, at most, we find a fractional deviation of $4\%$.
This behavior suggests a reliable way to measure $i$ from $\alpha_h$.

The behavior of $\overline{b_h}$ is more variable in comparison.
The top panel of \autoref{fig:ivq} shows the relationship between $\overline{b_h}$, $q$, and $i$ for an inner shadow with $s{=}0\degree$.
We find that $\overline{b_h}$ grows non-linearly with decreasing charge and increasing inclination.
Similarly, the bottom panel of \autoref{fig:ivq} shows the relationship between $\overline{b_h}$, $q$, and $s$ at $i{=}0\degree$, where we see that $\overline{b_h}$ grows at a near linear rate with ${\sim}\sqrt{\Delta q^2+\Delta s^2}$ for most values of $q$.

Next, \autoref{fig:constraints} shows how various assumptions can affect the ability of accurate measurement of $\overline{b_h}$ to constrain different regions of parameter space.
The top panel shows how the assumption of a known $s$ constrains a region of space between $q$ and $i$.
In this example, a perfect measurement of $\overline{b_h}$ constrains $q$ between the values of $[0.16, 0.75]$.
However, the constraint region is heavily dependent on measurement quality (red band).
We also find that an additional measurement of the shadow size (blue band) could further constrain $q$ and $i$, even if imperfect.

The bottom panel of \autoref{fig:constraints} shows the related problem of constraining $q$ and $s$ from a measurement of $\overline{b_h}$, assuming a known $i$. 
Because $\overline{b_h}$ increases radially outward, uncertainties of $1\%$ and $5\%$ trace bands of a near constant width.
We note that even though the critical curve is insensitive to $s$, making a measurement of $b_c$ along with $\overline{b_h}$ provides additional information that can be used to further constrain both $q$ and $s$.

\subsection{\m87-like system: Measurements of Charge and Mass}

Here we study the ability to constrain $q$ and $M/D$ from measurements of $\overline{b_h}$ and $b_c$, under the assumption of accurate measurements in $i$, but uncertainty in $s$.
Our results are summarized in \autoref{fig:stacked_plots}.
In general, dynamical models of black hole accretion often feature emission that is off the equatorial plane, altering inner shadow morphology \citep{Chang_2024, m87V}.
This scenario is relevant for studies of systems like \m87 where the mass of the system may not be well constrained, but where jet measurements can be used to infer orientation \citep{Walker,m87V}.

The top panel of \autoref{fig:stacked_plots} shows how a constraint can be made on $q$ and $M/D$ from measurements of $\overline{b_h}$ and $b_c$ if we have assumed the correct value of $s{=}0\degree$. 
The behavior of $\overline{b_h}$ and $b_c$ mirrors the results of \cite{Chael_2021}, which uses similar measurements to constrain mass and spin.
In contrast, the center and bottom panels illustrate the confounding effect that false assumptions on $s$ can have on parameter inference by introducing systematic errors.
These two behaviors highlight the importance of considering variations in emission geometry when performing measurements from black hole images. 

\section{Conclusion}

We study the ability to constrain an observer's viewing inclination $i$, emission co-latitude $s$, black hole charge $q$, and black hole mass-to-distance ratio $M/D$ from measurements of the shadows and inner-shadows of Reissner-Nordström black holes.
We find that the inner-shadow asymmetry, $\alpha_h$, has a strong dependence on $i$, but is mostly insensitive to changes in $q$ and $s$.
$\alpha_h$ thus serves as a robust probe of observer inclination which is relevant to observations of \sgra where $M/D$ is known, but $i$ is not.

We find that the average size of the inner-shadow $\overline{b_h}$ is also sensitive to $i$, but in contrast to $\alpha_h$, is strongly dependent on charge and emission geometry.
We show that this sensitivity allows independent radii measurements of the shadow and inner-shadow to constrain $M/D$ and $q$ if $i$ is known; a scenario that is relevant for observations on \m87.
In this regard, the behavior of the inner-shadow to constrain $q$ of a Reissner-Nordström is similar to its ability to constrain the spin of a Kerr black hole.
However, we show that $s$ introduces systematic error in measurements of $q$ and $M/D$ if incorrectly assumed.
These results emphasize the importance of incorporating varying emission geometry into models of black hole accretion flows \citep[eg.][]{Zhang_2024,Chang_2024}.

\begin{acknowledgments}
\section*{Acknowledgements}
We thank Richard Anantua and Prashant Kocherlakota for their insightful discussions.
Julien Kearns was
supported by the NSBP/SAO EHT Scholars program.
Support for this work was provided by the NSF through grant AST-1952099.
We acknowledge financial support from the National Science Foundation (AST-2307887).
This publication is funded in part by the Gordon and Betty Moore Foundation, Grant GBMF-12987. 
This work was supported by the Black Hole Initiative, which is funded by grants from the John Templeton Foundation (Grant \#62286) and the Gordon and Betty Moore Foundation
(Grant GBMF-8273)---although the opinions expressed in this work are those of the author(s) and do not necessarily reflect the views of these Foundations. 
\end{acknowledgments}

\bibliographystyle{apsrev4-2}
\bibliography{Refs,EHTRefs}
\end{document}